\documentclass[12pt]{iopart}

\newcommand{\eq}[1]{\begin{equation}#1\end{equation}}
\newcommand{\dd}{\mathrm{d}}

\newcommand{\acosh}{\mathrm{acosh\,}}
\usepackage{amssymb}
\usepackage{bbold}
\usepackage{graphics}
\usepackage{graphicx}
\usepackage{psfrag}

\begin{document}

\title[The Quantum Exclusion Process]
{Crossover between ballistic and diffusive transport: The Quantum Exclusion Process}

\author{Viktor Eisler$^{1,2}$}

\address{$^1$ Niels Bohr Institute, University of Copenhagen, Blegdamsvej 17,
DK-2100 Copenhagen \O, Denmark \\
$^2$ Faculty of Physics, University of Vienna, Boltzmanngasse 5,
A-1090 Vienna, Austria}
\eads{\mailto{viktor.eisler@univie.ac.at}}

\begin{abstract}
We study the evolution of a system of free fermions in one dimension under the simultaneous effects of coherent tunneling
and stochastic Markovian noise. We identify a class of noise terms where a hierarchy of decoupled equations for the
correlation functions emerges. In the special case of incoherent, nearest-neighbour hopping the equation for the
two-point functions is solved explicitly. The Green's function for the particle density is obtained analytically and
a timescale is identified where a crossover from ballistic to diffusive behaviour takes place. The result can be
interpreted as a competition between the two types of conduction channels where diffusion dominates on large
timescales.

\end{abstract}

\section{Introduction}

Transport properties of one-dimensional quantum systems show intriguing features and have been
the topic of intensive research. Despite the efforts, some important aspects of the problem still lack a
conclusive answer.
The main open question is whether ballistic transport, characteristic of a number of integrable quantum 
systems at zero temperature, could survive thermal noise or rather a transition to diffusion is inevitable.
Paradigmatic integrable model systems include weakly interacting fermions or, in the context of spin systems,
the XXZ model where a number of analytical methods are available \cite{Giamarchibook}.
In spite of the long-standing conjecture that integrability protects the ballistic features of transport
at any finite temperature \cite{Castella95}, recent calculations have pointed out the existence of
a dominant diffusive transport channel in the gapless phase of the XXZ model at half filling \cite{Sirker09,Sirker11}.
The above results, along with a large number of different numerical and analytical works \cite{HHB07},
are based on linear response theory and the calculation of the conductivity through Kubo's formula.

A completely different approach to the transport problem is possible in the framework of open quantum systems
\cite{BPbook}. Under certain assumptions on the coupling to the environment, the time evolution can be cast
in the form of a quantum master equation for the system density matrix. In the context of spin chains
this approach was initiated in \cite{Wichterich07} by modeling the coupling of the chain at both ends to heat
baths of different temperatures. In the Markovian approximation, an analytical treatment of the problem was
presented \cite{Prosen08} by diagonalizing the time evolution operator for spin chains that can be mapped
into a quadratic fermionic system. In the steady state of the XX chain (respectively free fermions) a flat
magnetization profile emerges \cite{KP09}, suggesting a ballistic transport. This has been recently supported
by a calculation of a lower bound on the Drude weight for the critical XXZ chain \cite{Prosen11}.

The above approach is restricted to incoherent processes which either create or annihilate a particle.
Furthermore, most of the cases considered so far were restricted to boundary driving.
An interesting example, where particle-conserving dephasing noise is present in the bulk of an XX chain
was presented in \cite{Znidaric10}. The steady state was calculated analytically in a perturbation series
with respect to the boundary driving and some exact results show a diffusive, linear magnetization profile
for any finite value of the bulk noise. Similar results were found numerically for the XXZ chain
\cite{Znidaric09} as well as in the case when the noise term describes stochastic hopping \cite{TWV09}.

It must be emphasized, that most of the above mentioned examples considered only steady-state properties
and the few existing results on the time evolution are limited to numerical methods \cite{Clark09,PZ09}.
On the other hand, there exists a remarkable exact solution for the complete spectrum of the time evolution
operator of a quantum diffusion problem \cite{EG05,EG05prb}. This is, however, restricted to a single fermion
moving on a chain with coherent tunneling and subject to dephasing noise. Therefore, it would be favorable
to find examples where time evolution under particle-conserving stochastic noise can be exactly
tackled in the context of a genuine many-particle problem.
This is further motivated by the fact, that such models have been recently suggested for the description
of energy transfer in photosynthetic complexes and biomolecules \cite{Mohseni08,Plenio08}.

In the present paper we discuss the evolution equations of free fermions when the coherent tunneling
is supplemented by thermally activated, stochastic hopping. This choice is favored since in the infinite
temperature limit it reproduces the well-known symmetric simple exclusion process \cite{Liggettbook}.
An investigation of the evolution operator shows a remarkable hierarchy property of the correlation functions
which enables us to derive the exact equations of motion for the two-point correlations. These equations
turn out to be very similar to the case of the single particle quantum diffusion \cite{EG05} and can be
explicitly solved. A further analysis of the resulting expressions for the time-dependent density profile yields
an exact analytical form of the Green's function. This formula is then used to identify a well-defined timescale
which separates a ballistic transport regime for short times from the diffusive behaviour in the long time limit.
The effective diffusion constant diverges in the limit of vanishing noise, signaling the transition to the
pure ballistic regime.

In Section \ref{sec:me} we investigate the master equation for the density matrix and introduce a class of
stochastic processes where the structure considerably simplifies and a hierarchy of decoupled equations emerges.
The simple case where the stochastic terms describe symmetric nearest-neighbour exclusion is introduced in
Sec. \ref{sec:sqep} followed by the derivation of the master equations for the two-point correlations.
Section \ref{sec:sol} is devoted to the analytical solution of these equations. In Section \ref{sec:gf} we give an
explicit analytical form of the Green's function for the density profile and compare it with numerical results.
Our findings are discussed in the last Section \ref{sec:disc} and some details of the calculations are presented
in two Appendices.

\section{The master equation \label{sec:me}}

The coherent evolution of closed quantum systems is described by a unitary time evolution operator.
However, in any realistic situation one has an inevitable coupling to the environment which introduces
incoherent effects. Under certain assumptions, the description of these open quantum systems is
possible in terms of a quantum master equation involving the density matrix of the system \cite{BPbook}.
In many examples the dynamics of the environment has a much shorter time scale and it is reasonable
to make the Markovian approximation. Then the time evolution of the system density matrix $\rho$ is,
in general, given by a master equation of the Lindblad form \cite{Lindblad76}
\eq{
\frac{\dd}{\dd t}\rho = \mathcal{L}(\rho) =-i\left[H,\rho\right]+ 
\sum_\alpha \left(L_\alpha \rho L_\alpha^\dag - 
\frac{1}{2}\{ L_\alpha^\dag L_\alpha,\rho \} \right)
\label{eq:lme}}
where the symbol $\mathcal{L}$ refers to the Liouvillian. The first term in the equation corresponds to a
coherent time evolution according to Hamiltonian $H$, while the operators $L_\alpha$ describe different
sources of stochastic noise. 

We will focus on fermionic quantum systems defined on a one-dimensional chain of length $N$.
The degrees of freedom are described by the creation and annihilation operators $a_m$ and $a_m^\dag$
possessing canonical anticommutational relations $\left\{a_m,a_n^\dag\right\}=\delta_{m,n}$ and 
$\left\{a_m,a_n\right\}=0$ for $m,n=1,\dots,N$. 
We will be interested in model systems where the coherent evolution is given by a free fermionic Hamiltonian
that is quadratic in the creation and annihilation operators. However, without further assumptions on the
Lindblad operators $L_\alpha$, solving the master equation (\ref{eq:lme}) turns out to be a very difficult problem.

Recently a simple integrable example was given by Prosen \cite{Prosen08}. Here, the Lindblad operators were taken
to be an arbitrary \emph{linear combination} of the fermionic operators and in turn describe processes involving the creation
or the loss of a particle. The integrability of the master equation relies on the fact that the operators $L_\alpha$
appear quadratically in (\ref{eq:lme}) and induce a Gaussian time evolution operator. This is, however, no longer true
if one considers stochastic processes (e.g. simple exclusion) that conserve the number of particles. Then $L_\alpha$
involve \emph{quadratic terms} in the fermi operators and the Liouvillian is not any more diagonalizable by a canonical 
transformation. However, as it will be shown below, some particular choice for the incoherent terms leads to a considerable
simplification of the problem.

We will follow the formulation of Ref. \cite{Prosen08} and start by introducing $2N$ Majorana fermions
with the definition
\eq{c_{2m-1}=a_m+a_m^\dag \, , \qquad c_{2m}=i(a_m-a_m^\dag)}
that are Hermitian and satisfying the relations $\left\{ c_k , c_l \right\} =2 \delta_{k,l}$ for all $k,l=1,\dots, 2N$. 
In terms of these operators the most general quadratic Hamiltonian and Lindblad operators read
\eq{
H=\frac{i}{4}\sum_{k,l=1}^{2N}H_{kl} c_k c_l \, , \qquad
L_\alpha=\frac{i}{4}\sum_{k,l=1}^{2N}L_{\alpha,kl} c_k c_l
\label{eq:hlmatrix}}
where $H_{lk}=-H_{kl}$ is required by hermiticity of the Hamiltonian. Note, that this choice of $H$ and $L_\alpha$ corresponds to
coherent and incoherent processes that either conserve the particle number or involve pair creation and annihilation.

It turns out to be more convenient to work with observables instead of density operators. For this purpose
one can introduce a convenient basis and define the ordered strings of Majorana operators
\eq{\Gamma_{\underline{\nu}}=c_1^{\nu_1} \dots c_{2N}^{\nu_{2N}}, \qquad
\nu_i \in \left\{ 0,1 \right\}
\label{eq:npf}}
where $\underline{\nu} = (\nu_1,\dots,\nu_{2N})$ denotes the vector of the occupation numbers $\nu_i$
indicating whether the corresponding Majorana operator $c_i$ is present in the string $\Gamma_{\underline{\nu}}$.
These objects encode the different $n$-point correlation functions where the order is given by
$n=\sum_i \nu_i$. It is also useful to define \emph{superoperators} acting on these strings that create or annihilate
a Majorana operator at position $j$ as
\eq{
\hat{c}_j \Gamma_{\underline{\nu}} = \delta_{1,\nu_j}\pi_j\Gamma_{\underline{\nu}'}, \qquad
\hat{c}_j^{\dag} \Gamma_{\underline{\nu}} = \delta_{0,\nu_j}\pi_j\Gamma_{\underline{\nu}'}, \qquad
\nu'_i = \left\{
\begin{array}{ll}
1-\nu_i & i=j \\
\nu_i & i \ne j
\end{array} \right.
\label{eq:so}}
where the sign factor $\pi_j= \exp\left( i\pi \sum_{k=1}^{j-1}\nu_k\right)$ ensures that canonical anticommutational
relations $\{ \hat{c}_i , \hat{c}_j^{\dag} \} = \delta_{i,j}$ and $\left\{ \hat{c}_i , \hat{c}_j \right\}=0$ are satisfied.
Note, that these superoperators change the order of the correlation function from $n$ to $n \pm 1$.
The overall number of independent, ordered correlation functions is $2^{2N}$ which exactly
equals the number of components in the density matrix $\rho$. This indicates, that solving the master equation
(\ref{eq:lme}) for $\rho$ is equivalent with solving the complete set of master equations for the operators
$\Gamma_{\underline{\nu}}$ which are generated by the adjoint Liouvillian $\mathcal{L^\dag}$.

We are now ready to state the main result of this section. If the Lindblad operators satisfy $L_\alpha^\dag = L_\alpha$
for all $\alpha$ the Liouvillian takes the following simple form
\eq{
\mathcal{L^\dag} = -\sum_{k,l=1}^{2N} \tilde H_{kl} \hat{c}^{\dag}_k \hat{c}_l
+ \frac{1}{2} \sum_{i,j,k,l=1}^{2N} \sum_{\alpha}
L^T_{\alpha,ij} L_{\alpha,kl} \hat{c}^\dag_i\hat{c}^{\dag}_k \hat{c}_j\hat{c}_l 
\label{eq:liou}}
where $\tilde H_{kl}$ denote the elements of the matrix ${\bf \tilde H}={\bf H}+\frac{1}{2}\sum_\alpha{\bf L}^T_\alpha {\bf L}_\alpha$.
It is important to stress that the operator (\ref{eq:liou}) conserves the number of Majorana fermions and therefore
the length of the strings $\Gamma_{\underline{\nu}}$. In other words, a hierarchy of the $n$-point correlation functions
emerges since the time evolution under $\mathcal{L^\dag} $ does not mix strings of different length.
Note, that a similar hierarchy property was pointed out for the steady state of the driven XX chain with dephasing
\cite{Znidaric11}, which is now seen to generalize to the complete time evolution.
For the derivation of (\ref{eq:liou}) we refer to Appendix A.

Before fixing our model, it is worth investigating the general structure of Eq. (\ref{eq:liou}). The first term contains the
Hamiltonian evolution modified by a damping term which is in turn responsible for the decay of the $n$-point functions.
This quadratic term generates the Gaussian part of the time evolution and has exactly the same form as in case of linear
Lindblad operators \cite{Prosen08}. However, one has an additional ``interaction term'' which is of fourth order in the
superoperators and in general implies that time evolution is non-Gaussian. Although the Liouvillian cannot be
diagonalized in general as in \cite{Prosen08}, some simple choice of the stochastic terms could further simplify
(\ref{eq:liou}) and eventually lead to a tractable problem.

\section{The symmetric quantum exclusion process\label{sec:sqep}}

\begin{figure}[htb]
\center
\includegraphics[scale=0.5]{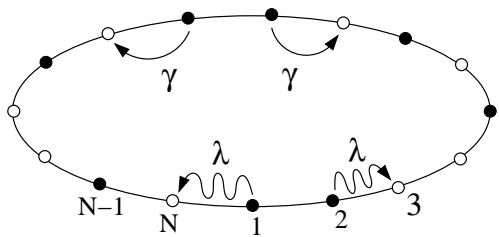}
\caption{Symmetric quantum exclusion process on a ring of size $N$. Particles can move to
unoccupied nearest neighbour sites by coherent tunneling (curled arrows) or stochastic hopping
(simple arrows) with corresponding rates $\lambda$ and $\gamma$.}
\label{fig:geom}
\end{figure}

After setting up the general formalism, we proceed to specifying the concrete model.
Our focus is to define a dynamics which interpolates between the extreme cases of quantum coherent tunneling,
described by the tight-binding Hamiltonian, and classical stochastic hopping, described by the
symmetric simple exclusion process \cite{Liggettbook}. Such a model has been introduced recently \cite{TWV09}
and investigated numerically from the perspective of steady state properties. In order to get a nontrivial steady state,
boundary injection and ejection was introduced into the dynamics. Since here we are interested in the complete
time-dependent solution, it turns out to be more convenient to take periodic boundary conditions.
The geometry and the update rules are sketched in Figure \ref{fig:geom}.

%

The coherent evolution is described by the Hamiltonian
\eq{
H = -\lambda \sum_{j=1}^{N} (a_j^\dag a_{j+1}+a_{j+1}^\dag a_j)
\label{eq:ham}}
where $\lambda$ is the tunneling rate. The incoherent hopping is generated by the Lindblad operators
\eq{
L_{Lj}=\sqrt{\gamma} a_j^\dag a_{j+1} \, , \qquad
L_{Rj}=\sqrt{\gamma} a_{j+1}^\dag a_j
\label{eq:lop}}
where $L_{Lj}$ and $L_{Rj}$ with $j=1,\dots,N$ describe hopping to the left and right neighbour sites with the same rate $\gamma$.
Although these Lindblad operators are not Hermitian, one can introduce new ones with a simple unitary transformation
\eq{
L_{2j-1} = \frac{1}{\sqrt{2}}\left( L_{Lj}+L_{Rj} \right) \, , \qquad
L_{2j} = \frac{i}{\sqrt{2}}\left( L_{Lj}-L_{Rj} \right)
\label{eq:lop2}}
which now satisfy the conditions $L_{\alpha}=L_{\alpha}^{\dag}$ for every $\alpha=1,\dots,2N$.
Since the master equation (\ref{eq:lme}) is invariant under such unitary transformations, one can now apply
the results of the previous section. Note, that this step requires symmetric hopping.

The matrix elements of $H$ in the Majorana basis can be written in a block matrix notation as
\eq{
H_{mn} = -\lambda
\left(\begin{array}{cc}
0 & -1 \\
1 & 0
\end{array}\right)
\otimes (\delta_{m,n-1} + \delta_{m,n+1})
\label{eq:hbm}}
where the indices take the values $m,n=1,\dots,N$. Similarly, one has
\eq{
\eqalign{
L_{2j-1,mn}  = \sqrt{\frac{\gamma}{2}} 
\left(\begin{array}{cc}
0 & -1 \\
1 & 0
\end{array}\right)
\otimes (\delta_{m,j}\delta_{n,j+1} + \delta_{m,j+1}\delta_{n,j})
\\
L_{2j,mn} = \sqrt{\frac{\gamma}{2}}
\left(\begin{array}{cc}
1 & 0 \\
0 & 1
\end{array}\right)
\otimes (\delta_{m,j}\delta_{n,j+1} - \delta_{m,j+1}\delta_{n,j})
\\
\frac{1}{2}\sum_{\alpha=1}^{2N} {\bf L}^T_{\alpha}{\bf L}_{\alpha} = \gamma
\left(\begin{array}{cc}
1 & 0 \\
0 & 1
\end{array}\right)
\otimes \delta_{m,n}
}\label{eq:lbm}}
The last matrix is diagonal and describes a pure damping term. Moreover, all the other matrices have a simple block-tridiagonal form,
however some non-diagonal entries appear. Therefore, we perform the following
canonical transformation of the superoperators
\eq{
\hat{a}_m = \frac{1}{\sqrt{2}}\left(\hat{c}_{2m-1} - i\hat{c}_{2m}\right) \, , \qquad
\hat{b}_m = \frac{1}{\sqrt{2}}\left(\hat{c}_{2m-1} + i\hat{c}_{2m}\right) 
}
which will diagonalize the $2 \times 2$ matrices in $H$ and $L_{2j-1}$ with eigenvalues $\pm i$. Note, that the effective action of the
superoperators $\hat{a}_{m}$ and $\hat{b}_{m}$ is to remove the fermionic operators $a_m$ and $a_m^\dag$ from a string.

Substituting into (\ref{eq:liou}) a simple calculation yields the Liouvillian of the symmetric quantum exclusion process in the form
$\mathcal{L^\dag} = \mathcal{L^\dag}_{a} + \mathcal{L^\dag}_{b} + \mathcal{L^\dag}_{ab}$ where
\eq{
\eqalign{
\mathcal{L^\dag}_{a} &=
\sum_{m} \left[ i\lambda \left( \hat{a}_m^\dag \hat{a}_{m+1}  + \hat{a}_{m+1}^\dag \hat{a}_{m} \right)
-\gamma \hat{a}_m^\dag \hat{a}_{m}
+\gamma \hat{a}_m^\dag \hat{a}_{m} \hat{a}_{m+1}^\dag \hat{a}_{m+1} \right],
\\
\mathcal{L^\dag}_{b} &=
\sum_{m} \left[ -i\lambda\left( \hat{b}_m^\dag \hat{b}_{m+1} + \hat{b}_{m+1}^\dag \hat{b}_{m} \right)
-\gamma \hat{b}_m^\dag \hat{b}_{m} 
+\gamma \hat{b}_m^\dag \hat{b}_{m} \hat{b}_{m+1}^\dag \hat{b}_{m+1} \right],
\\
\mathcal{L^\dag}_{ab} &=
\sum_{m} \gamma \left( \hat{a}_m^\dag \hat{a}_{m+1} \hat{b}_m^\dag \hat{b}_{m+1} +
\hat{b}_{m+1}^\dag \hat{b}_{m} \hat{a}_{m+1}^\dag \hat{a}_{m}\right).
}
\label{eq:liouqep}}
The operator $\mathcal{L^\dag}_{a}$ is therefore formally equivalent to the one describing the motion of a 1D fermionic system
in a chemical potential and with repulsive nearest neighbour interactions. Note, however, that one has an imaginary hopping
amplitude $i\lambda$. The term $\mathcal{L^\dag}_{b}$ describes a second species of fermions where the only difference
is in the hopping amplitude $-i\lambda$. The last term translates into a special interaction between the two types of fermions,
giving a penalty term for simultaneous hopping. 

The time evolution of the correlators of the quantum exclusion process is obtained by systematically applying (\ref{eq:liouqep})
to the various strings of operators, which is demonstrated below on the simplest examples.

\subsection{One-point correlations}

The simplest string consists of only one fermion operator. The contributions of the interaction terms then vanish and one has
\eq{
\frac{\dd}{\dd t} a_m =\mathcal{L^\dag}_{a}(a_m) = i\lambda \left( a_{m-1} + a_{m+1}\right) -\gamma a_{m}
\label{eq:evola}}
while the equation for $a_m^{\dag}$ follows simply by hermitian conjugation. Using the Fourier transformed operators $a_q$,
this equation can be solved as
\eq{
a_q = e^{\varepsilon t} a_q (0), \qquad
\varepsilon = 2i\lambda \cos q - \gamma t
}
where $q=2\pi n/N$ for $n=1,\dots,N$. Since most of the physically interesting initial states will give a vanishing expectation value
$\langle a_q(0) \rangle$, we move forward to analyze the first nontrivial correlation functions.

\subsection{Two-point correlations}

The most basic physical quantities, such as the particle density, are encoded in the two-point correlations.
The independent correlators are chosen as $a_m^{\dag} a_n$ for $m \le n$ and $a_m^{\dag} a_n^{\dag}$
for $m<n$, while all the other combinations are related by complex conjugation and the commutational relations.
Instead of dealing with operators, one can already take the expectation values with respect to
some arbitrary initial state. The equation for $G_{m,n} = \langle a_m^{\dag} a_n \rangle$ then reads
\eq{
\eqalign{
\frac{\dd}{\dd t} G_{m,n} = &-i \lambda
\left( G_{m-1,n} + G_{m+1,n} - G_{m,n-1} - G_{m,n+1} \right) - 2 \gamma G_{m,n} \\
&+\gamma \delta_{m,n} \left(G_{m-1,m-1} + G_{m+1,m+1} \right)}
\label{eq:evolg}}
where the contribution in the second line is generated by $\mathcal{L^\dag}_{ab}$.
Similarly, the evolution of the pair-creation expectation values $F_{m,n} = \langle a_m^{\dag} a_n^{\dag} \rangle$  is given by
\eq{
\eqalign{
\frac{\dd}{\dd t} F_{m,n} = &-i \lambda
\left( F_{m-1,n} + F_{m+1,n} + F_{m,n-1} + F_{m,n+1} \right) - 2 \gamma F_{m,n} \\
&+\gamma \delta_{m,n-1} F_{m,m+1} - \gamma \delta_{m,n+1} F_{m-1,m}}
\label{eq:evolf}}
Since both (\ref{eq:evolg}) and (\ref{eq:evolf}) are linear, first order differential equations,
they can be formulated as a matrix eigenvalue problem. The interaction term of the Liouvillian
has only a localized contribution proportional to $\delta_{m,n}$ and $\delta_{m,n\pm 1}$, respectively,
and therefore the equations translate into a simple potential scattering problem.

\section{Solution of the scattering problem\label{sec:sol}}

The equation (\ref{eq:evolg}) for the particle-hole correlators has a similar form to the one appearing
in \cite{EG05} as the equation of motion for a single electron with dephasing noise.
The solution of the scattering problem is presented in detail for $G_{m,n}$ along the lines of Ref. \cite{EG05}
and we also briefly comment on the analogous treatment for $F_{m,n}$.

\subsection{Particle-hole correlations \label{sec:scpg}}

The linearity of (\ref{eq:evolg}) allows one to separate the time dependent part as
$G_{m,n}(t)=G_{m,n}e^{\varepsilon t}$ with the energy eigenvalue $\varepsilon$.
Translational invariance implies, that the eigenvectors obey $G_{m,n}=e^{iqm}G_{0,n-m}$
where the allowed values of the wavenumber are $q_j=\frac{2\pi}{N}j$ with $j=1,\dots,N$.
Introducing the relative coordinate $l=n-m$ and substituting $G_{0,l}=i^{-l}e^{i\frac{q}{2}l}g_l$
we get the simpler equation
\eq{
\varepsilon g_l = 2i\lambda\sin\frac{q}{2}(g_{l-1}+g_{l+1}) - 2\gamma g_l
+ 2\gamma \cos q \, g_l \, \delta_{l,0}
\label{eq:gl}}
where $0\le l \le N-1$ and the boundary conditions imply $g_{N}=i^Ne^{-i\frac q 2 N} g_0$.
One recognizes, that the structure of Eq. (\ref{eq:gl}) for the amplitude $g_l$
is the same as the one describing the motion of a free particle with a potential scattering
center localized at $l=0$.  This is solved by using the ansatz
\eq{
g_l = A e^{i\theta l} + B e^{-i\theta l}
\label{eq:scp}}
Substituting into (\ref{eq:gl}) one obtains $N-2$ identical equations that fix the
energy eigenvalue as
\eq{
\varepsilon_{q \theta} = 4i \lambda \sin \frac{q}{2}\cos\theta- 2\gamma
\label{eq:eps}}
where $q$ and $\theta$ are explicitly used to index the eigenvalues.
The remaining two equations for $l=0$ and $l=N-1$ determine the quantization of the
second quantum number $\theta$ and fix the value $B/A$ of the scattering phase.
The solution of these equations follows exactly along the lines of Ref. \cite{EG05} and is
summarized in Appendix B. In turn, one finds that the eigenmodes can be written in the
following Bethe ansatz form
\eq{
G^{q,\theta}_{m,n}=
z_1^{m} z_2^{n} + S_{21} (-z_2)^{m} (-z_1)^{n}
\label{eq:bag}}
where $S_{21}=B/A$ and we used the notation
\eq{
z_1 = e^{i(q/2-\theta+\pi/2)}, \qquad
z_2 = e^{i(q/2+\theta-\pi/2)} \, .
\label{eq:z12}}

\begin{figure}[t]
\center
\includegraphics[scale=0.6]{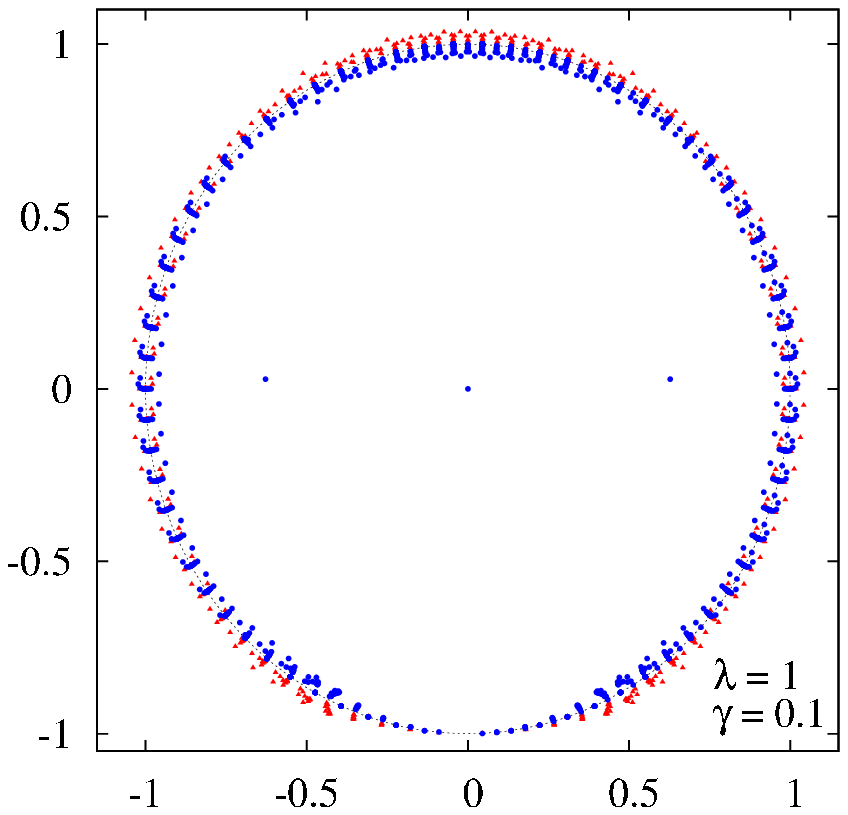}
\includegraphics[scale=0.6]{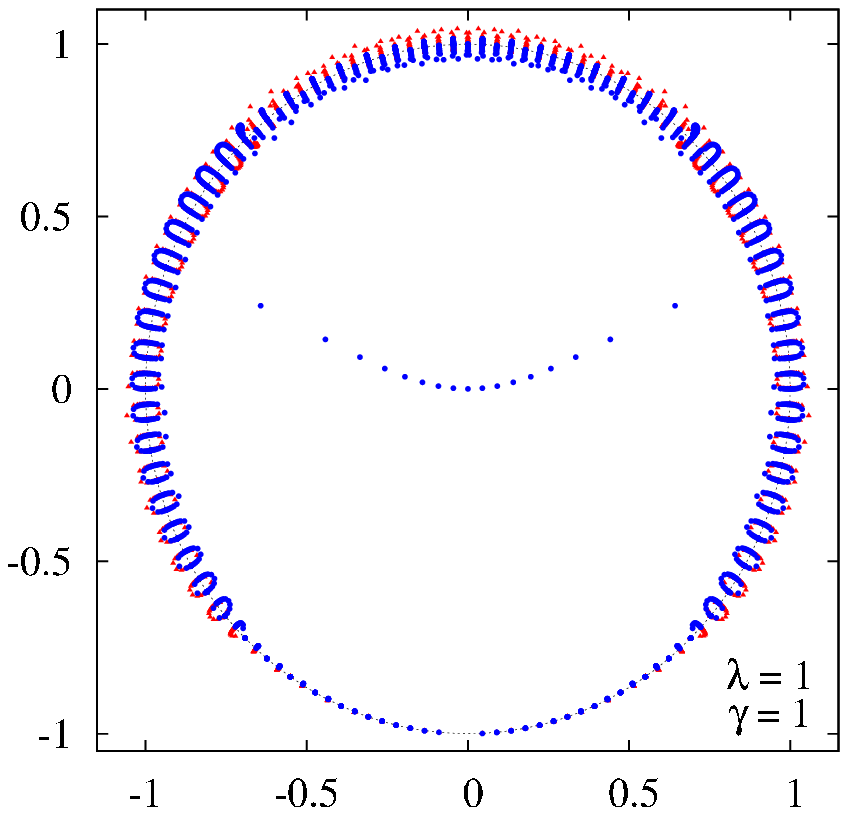}
\includegraphics[scale=0.6]{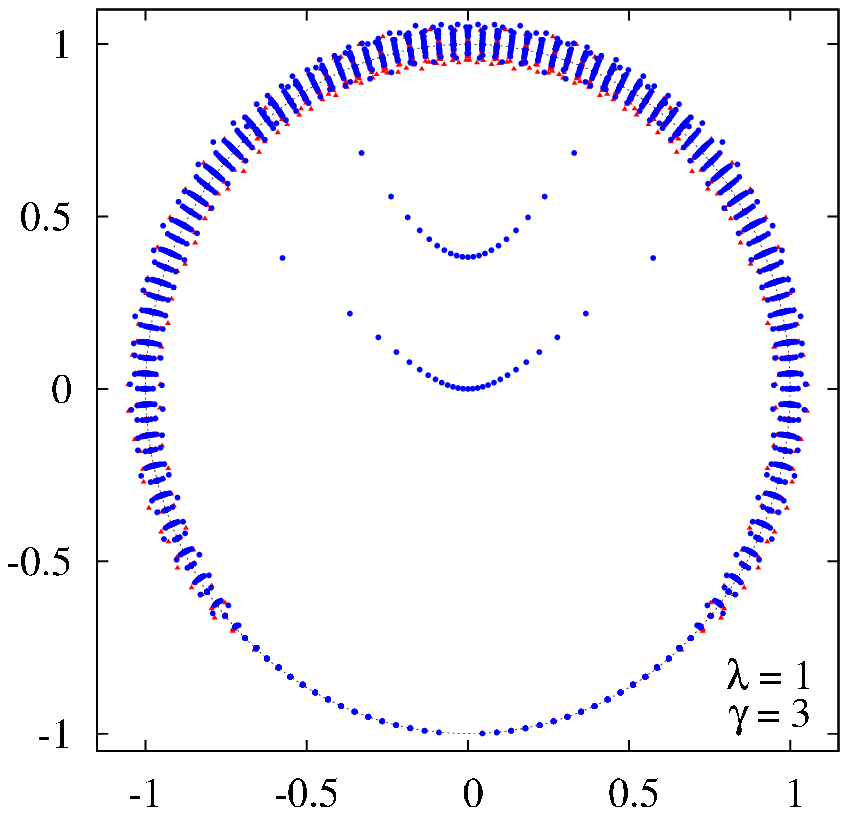}
\caption{Allowed values of the parameters $z_1$ (red dots) and $z_2$ (blue dots) on the complex
plane for $N=70$. The tunneling rate is $\lambda=1$ and the hopping rate takes the values
$\gamma=0.1$ (left), $\gamma=1$ (middle) and $\gamma=3$ (right). The $z_1$ solutions
of the diffusive branch lie further outside of the unit circle and are not shown.}
\label{fig:z1z2}
\end{figure}

The allowed wavenumbers $\theta$ for finite $N$ can only be determined numerically 
by solving Eq. (\ref{eq:beq}). The resulting $z_1$ and $z_2$ are shown in Fig. \ref{fig:z1z2}
for $N=70$. In spite of the nonvanishing imaginary parts of $\theta$, one clearly sees
a condensation around the unit circle, corresponding to the eigenvalue families $\theta_{I}$
and $\theta_{II}$. However, the third family $\theta_{d}$ leads to a separate branch of solutions
$z_1$ and $z_2$ on the complex plane, lying either inside or outside (not shown in figure) of the
unit circle. These solutions correspond to \emph{purely real} energy eigenvalues
\eq{
\varepsilon_d = 2\gamma \sqrt{\cos^2 q - 4\frac{\lambda^2}{\gamma^2}\sin^2 \frac{q}{2}} - 2\gamma \, .
\label{eq:epsd}}
It is obvious from Fig. \ref{fig:z1z2} that their number increases for larger values of the stochastic noise
$\gamma$ and the solution $z_2=0$ in the center of the circle, corresponding to the steady state 
$\varepsilon_d =0$, is always present. These eigenvalues are, in turn, responsible for the emergence of
diffusion in the time evolution. It is intuitively seen by expanding (\ref{eq:epsd}) around the steady
state $q=0$ which gives
\eq{
\varepsilon_d \approx -D q^2, \qquad D = \gamma + \frac{\lambda^2}{\gamma}
\label{eq:epsd0}}
and therefore yields the usual form of a diffusive dispersion relation with diffusion coefficient $D$.
For the detailed discussion of the evolution of the particle density we refer to the next section.

%

In order to obtain the general solution matching to a given initial condition one has to first construct
the left eigenvectors and prove the orthonormality relation. One starts by noticing that the spectrum in
Eq. (\ref{eq:eps}) is degenerate for the pair of solutions $q \to 2\pi - q$ while the complex phases transform
as $z_1 \to (-z_2)^{-1}$ and $z_2 \to (-z_1)^{-1}$ under this change. Therefore, we write the left eigenvector as
\eq{
\bar{G}^{q,\theta}_{m,n}=
z_1^{-m} z_2^{-n} + S_{21}^{-1} (-z_2)^{-m} (-z_1)^{-n}
\label{eq:bagl}}
In order to obtain an orthonormal system one has to extend the eigenvectors to index pairs where $m>n$.
Since the solutions of equation (\ref{eq:gl}) are either even $g_{-l}=g_{l}$ or odd $g_{-l}=-g_{l}$, the right eigenvectors
must satisfy the symmetry property $G ^{q,\theta}_{n,m} = \pm (-1)^{m+n} G ^{q,\theta}_{m,n}$ and the same holds for
the left eigenvectors. Note, that the antisymmetric eigenvectors, insensitive to the value of the scattering
potential, are exactly the ones in the family $\theta_{II}$ (see Appendix B). Introducing the notation $\langle q',\theta' |$
for the left and $ |q,\theta\rangle$ for the right eigenvectors, a lengthy but simple calculation shows
\eq{
\langle q',\theta' |q,\theta\rangle = \sum_{m,n}\bar{G}^{q',\theta'}_{m,n}G^{q,\theta}_{m,n}=
\delta_{q,q'}\delta_{\theta,\theta'}N_{q \theta}
\label{eq:orth}}
where the sum goes over all the indices and the complex normalization factor reads
\eq{
N_{q\theta} = 2N(N+i\Delta), \qquad
\Delta = -\frac{2 \beta \cos \theta}{1-\beta^2\sin^2\theta}
\label{eq:nqt}}
where the parameter $\beta$ is defined in (\ref{eq:beta}). Finally, the complete solution can be written in the form
\eq{
|G(t)\rangle = \sum_{q,\theta} c_{q\theta} |q,\theta\rangle e^{\varepsilon_{q\theta}t}, \qquad
c_{q\theta}= N^{-1}_{q \theta}\langle q,\theta|G(0)\rangle
\label{eq:gt}}
where the constants $c_{q\theta}$ are determined through the initial condition $|G(0)\rangle$.

To conclude this section, we show that the inverse of the complex normalization constant $N^{-1}_{q\theta}$
can be interpreted as the density of states in the wavenumber space. First we note, that the solutions $\theta$
have a nonvanishing imaginary part which depends on the real part and thus also changes when going from
one solution to the next. Using the notation of Appendix B one has $\dd\theta = \dd \theta_0 + i\,\dd \delta$
and by differentiating (\ref{eq:th1<}) one finds $\dd \theta = 2\pi (N+i\Delta)^{-1}$. Therefore one has
$N^{-1}_{q\theta}=\dd q \, \dd \theta /8\pi^2$ which, up to a factor, indeed corresponds to the level density.

\subsection{Particle-particle correlations}

For completeness, we also remark on the solution of the pair-creation probabilities $F_{m,n}(t)$.
In principle, one has to follow the same steps as in case of the particle-hole correlations.
The time dependence is again of the form $F_{m,n}(t) = F_{m,n}e^{\tilde\varepsilon t}$ and the complex
parameters now have to be defined as $\tilde z_1 = e^{i(q/2-\theta)}$ and $\tilde z_2 = e^{i(q/2+\theta)}$,
respectively. It is easy to check that the ansatz
\eq{
F_{m,n}^{q\theta} = 
\tilde z_1^{m} \tilde z_2^{n} + \tilde{S}_{21} \tilde z_2^{m} \tilde z_1^{n}
\label{eq:baf}}
solves (\ref{eq:bag}) if the energy eigenvalue and the scattering phase satisfy
\eq{
\tilde\varepsilon_{q\theta} = -4i\lambda \cos\frac{q}{2}\cos\theta - 2\gamma , \qquad
\tilde{S}_{21} = - \frac{\tilde z_1 \tilde z_2 + 1 - i \frac{\gamma}{\lambda}\tilde z_2}
{\tilde z_1 \tilde z_2 + 1 - i \frac{\gamma}{\lambda}\tilde z_1} \, .
\label{eq:spf}}
The allowed values of the wavenumber $\theta$ is again determined by a corresponding
Bethe equation which we do not discuss in detail. Finally, one should remark that the
scattering phase (\ref{eq:spf}) has a very similar structure to the one appearing in the
Bethe ansatz solution of the simple exclusion process \cite{Schutz97}.

\section{Green function of the density\label{sec:gf}}

In this section we will focus on the time evolution of the particle density $G_{m,m}(t)$. 
The essential quantity to be determined is the Green's function $\mathcal{G}(m-k,t)$,
describing the evolution of a single particle localized at site $k$ for $t=0$.
The initial condition then reads
\eq{
G_{m,n}(0)= \delta_{m,k}\delta_{n,k}
\label{eq:gfic}}
and the scalar products with the left eigenvectors yield
\eq{
\langle q,\theta|G(0)\rangle = (1+S_{21}^{-1}) \, e^{-iqk} \, .
\label{eq:gfscp}}
Therefore, the family $\theta_{II}$ where $S_{21}=-1$ does not enter the solution.
Using the form (\ref{eq:scp1}) of the scattering phase the evolution of the density reads
\eq{
\mathcal{G}(m-k,t) =\sum_{q,\theta} N_{q\theta}^{-1}
\frac{-4\beta^2\sin^2\theta}{1-\beta^2\sin^2\theta}\,
e^{iq(m-k)}e^{\varepsilon_{q\theta} t}
\label{eq:gfsum}}
where the sum has to be taken over the families $\theta_{I}$ and $\theta_{d}$.
As remarked at the end of section \ref{sec:scpg}, the factor $N_{q\theta}^{-1}$ naturally translates
into the density of states in the $N \to \infty$ limit. The diffusive eigenvalues satisfy
$\beta\sin\theta_d \to 1$ but at the same time have a vanishing density $N_{q\theta}^{-1} \to 0$
in the $\theta$-space thus the product of the first two factors in (\ref{eq:gfsum}) is finite
and can be evaluated using (\ref{eq:nqt}). Hence, the Green's function is a sum of two separate
contributions
\eq{
\eqalign{
\mathcal{G}(x,t) &=
\int_{|\beta|<1} \frac{\dd q}{2\pi}e^{iqx}\frac{1}{\sqrt{1-\beta^2}} e^{\varepsilon_d t} \\
& - \int_{0}^{2\pi} \frac{\dd q}{2\pi}e^{iqx}
\int_{0}^{\pi}\frac{\dd \theta}{\pi}
\frac{\beta^2\sin^2\theta}{1-\beta^2\sin^2\theta} e^{\varepsilon _{q\theta} t}}
\label{eq:gfxint}}
where we introduced $x=m-k$. The first integral goes only over $q$ values satisfying $|\beta|<1$
and the $\theta$-integral has a pole for $|\beta|>1$, corresponding to the branch cut in the
$\theta_{I}$ eigenvalues. This suggests writing
\eq{
\mathcal{G}(x,t) = \int_0^{2\pi} \frac{\dd q}{2\pi}\mathcal{G}(q,t)e^{iqx}, \qquad
\mathcal{G}(q,t) = \left\{
\begin{array}{ll}
\mathcal{G}^{<}(q,t) \, , & \mbox{if } |\beta| < 1\\
\mathcal{G}^{>}(q,t) \, , & \mbox{if } |\beta| > 1
\end{array} \right.
\label{eq:gfx}}
where the Fourier transformed Green's function has to be defined piecewise. It must be emphasized,
however, that $\mathcal{G}(q,t)$ turns out to be a smooth and continuous function and its different forms on
both sides of $|\beta| = 1$ are connected by analytical continuation.

In order to obtain $\mathcal{G}^{<}(q,t)$, it is first useful to introduce the parameter
$\alpha = 2\gamma t \cos q$, in terms of which one has
$\varepsilon_d = \alpha \sqrt{1-\beta^2}-2\gamma t$ and 
$\varepsilon_{q\theta}=\alpha\beta\cos\theta-2\gamma t$. For $|\beta|<1$ the denominator
of the $\theta$-integrand can be expanded in a geometric series. The resulting integrals can be
carried out \cite{GRbook} and yield
\eq{
\mathcal{G}^{<}(q,t) = \left[\frac{e^{\alpha \sqrt{1-\beta^2}}}{\sqrt{1-\beta^2}}
-\sum_{n=1}^{\infty}\left(\frac{\beta}{\alpha}\right)^n J_n(\alpha\beta)\, (2n-1)!!\right] e^{-2\gamma t} \, .
\label{eq:gfq<}}
The above functional form is obviously only valid for $\beta <1$. In spite of the apparent singularity
of the first term for $\beta \to 1$, the sum also becomes divergent for these values and, since it enters
with the minus sign, regularizes the expression. In order to see this one can expand the exponential
in terms of its variables $\alpha$ and $\beta$ and further use a series representation of the Bessel
function \cite{GRbook}
\eq{
J_n(\alpha\beta)=\left(\frac{\alpha\beta}{2}\right)^n
\sum_{k=0}^{\infty}\frac{(-1)^k}{k! (n+k)!} \left(\frac{\alpha\beta}{2}\right)^{2k}
\label{eq:jab}}
which then yields $\mathcal{G}^{<} (q,t)$ as a double infinite sum. One can see that many terms cancel
out and, on one hand, one obtains the first term of (\ref{eq:gfq<}) with the exponential replaced by a
$\sinh$ function. Additionally, one has a regularized series of the same expression with $\cosh$,
where one of the sums is cut at a finite order. With a proper reordering of the latter terms and using
again (\ref{eq:jab}) one can rewrite it as a single sum involving Bessel functions. Finally, the term with
the hyperbolic sine can be analytically continued to $|\beta|>1$ and one arrives to the following
expression
\eq{
\mathcal{G}^{>}(q,t)=
\left[\frac{\sin\left(\alpha\sqrt{\beta^2-1}\right)}{\sqrt{\beta^2-1}}+
\sum_{n=0}^{\infty}\left(\frac{\alpha}{\beta}\right)^n
\frac{J_n(\alpha\beta)}{(2n-1)!!}\right]e^{-2\gamma t} \, .
\label{eq:gfq>}}
Note, that the first term is just the pole contribution of the integral (\ref{eq:gfxint}). Since the result
(\ref{eq:gfq>}) is obtained by analytic continuation rather then evaluating (\ref{eq:gfxint}) for $|\beta|>1$,
the continuity of the function $\mathcal{G}(q,t)$ is guaranteed.

The final Fourier transform (\ref{eq:gfx}) which yields the Green's function in coordinate space
in general has to be evaluated numerically. However, on some well defined timescales one finds simple
approximations to $\mathcal{G}(q,t)$ and the integral can be carried out explicitly.

%
\begin{figure}[htb]
\center
\psfrag{G(q,t)}[][][.6]{$\mathcal{G}(q,t)$}
\includegraphics[scale=0.6]{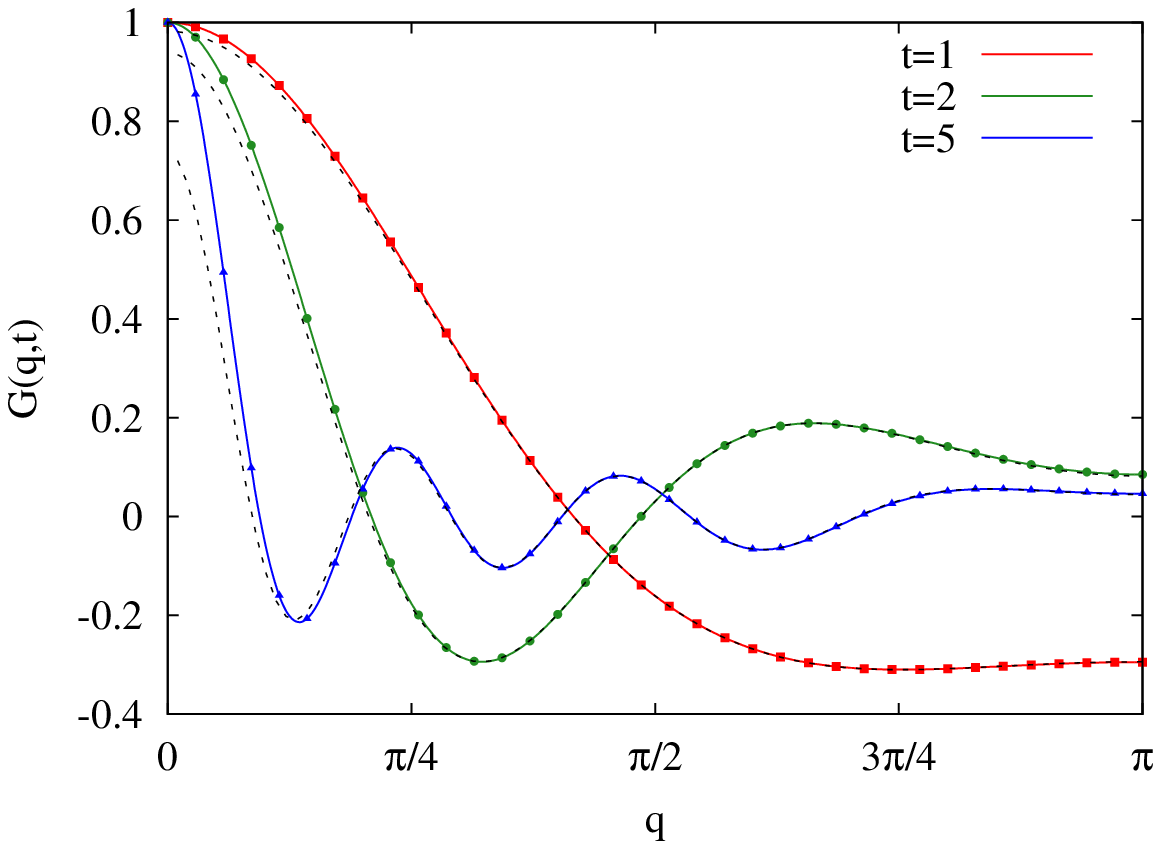}
\includegraphics[scale=0.6]{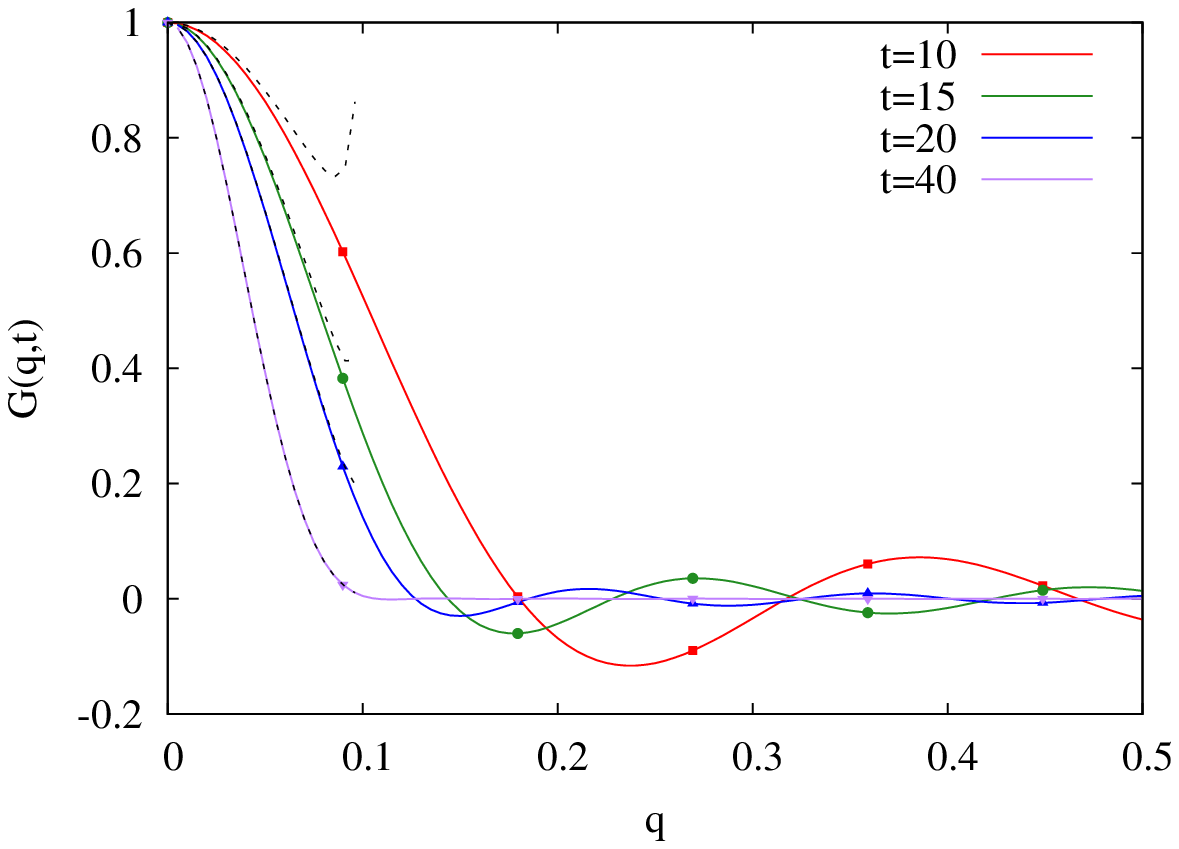}
\caption{Fourier transform $\mathcal{G}(q,t)$ of the Green's function for short (left) and long times (right)
with parameter values $\lambda=1$ and $\gamma=0.1$. Dashed lines correspond to the approximations
(\ref{eq:gfqst}) and (\ref{eq:gfqlt}), respectively. Points represent the exact solution for finite $N=70$
by evaluating the sum over $\theta$ in (\ref{eq:gfsum}).}
\label{fig:gfq}
\end{figure}

\subsection*{Short time behaviour}

On short timescales compared to the stochastic hopping rate, $\gamma t \ll 1$, one finds that for all wavenumbers
satisfying the condition $\alpha \ll \beta$ only the $n=0$ term in the sum of Eq. (\ref{eq:gfq>}) contributes
significantly and one has
\eq{
\mathcal{G}(q,t) \approx \left[\frac{\sin(\alpha\beta)}{\beta} + J_{0}(\alpha\beta) \right] e^{-2\gamma t} \, .
\label{eq:gfqst}}
The left hand side of Figure \ref{fig:gfq} shows that (\ref{eq:gfqst}) indeed gives a very good approximation
of $\mathcal{G}(q,t)$ on short timescales. However, as time increases the deviation becomes large around
$q \approx 0$ where $|\beta|<1$ and one cannot use the representation $\mathcal{G}^{>}(q,t)$.

The arguments in (\ref{eq:gfqst}) can be rewritten as $\alpha\beta=4\lambda t\sin \frac{q}{2}$.
By further assuming $\gamma \ll \lambda$, the first term can be neglected since it is multiplied by an additional
factor of $\beta^{-1} \sim \gamma/\lambda$. The remaining term can now be explicitly integrated as \cite{GRbook}
\eq{
\mathcal{G}(x,t) \approx \left[ J_{x}(2\lambda t)\right]^2 e^{-2\gamma t} \, .
\label{eq:gfxst}}
The resulting $\mathcal{G}(x,t)$ is simply the Green's function of the coherent case multiplied by an exponential
damping factor. Therefore, coherent effects are washed out on a timescale $\tau \sim \gamma^{-1}$.

\subsection*{Long time behaviour}

An other simple approximation can be obtained for large times $\gamma t \gg 1$. Then the sum in (\ref{eq:gfq>})
can be neglected unless $\alpha \gg \beta$ where one has to use the other representation $\mathcal{G}^{<}(q,t)$.
Since in (\ref{eq:gfq<}) the $n=0$ term is missing from the sum, one has
\eq{
\mathcal{G}(q,t) \approx \frac{e^{\varepsilon_d t}}{\sqrt{1-\beta^2}} \, .
\label{eq:gfqlt}}
The right hand side of Fig. \ref{fig:gfq} shows that for increasing times the oscillating tail of $\mathcal{G}(q,t)$
is suppressed and the approximation works well for the nonvanishing part at small wavenumbers.
Finally, using the expansion (\ref{eq:epsd0}) of $\varepsilon_d$ and noticing that for $\gamma t \gg 1$
the denominator in (\ref{eq:gfqlt}) can be set equal to $1$, one arrives to the simple result
\eq{
\mathcal{G}(x,t) \approx \frac{1}{\sqrt{4\pi Dt}}e^{-\frac{x^2}{4Dt}} \, .
\label{eq:gfxlt}}
Therefore, the long time behaviour of the evolution is always governed by diffusion.
Nevertheless, one sees from the expression of the diffusion constant (\ref{eq:epsd0}) that the speed of
diffusion can be significantly enhanced and actually diverges as $\gamma \to 0$, signaling the transition to
pure ballistic transport.

\vspace{1cm}

\begin{figure}[htb]
\center
\psfrag{G(x,t)}[][][.6]{$\mathcal{G}(x,t)$}
\includegraphics[scale=0.6]{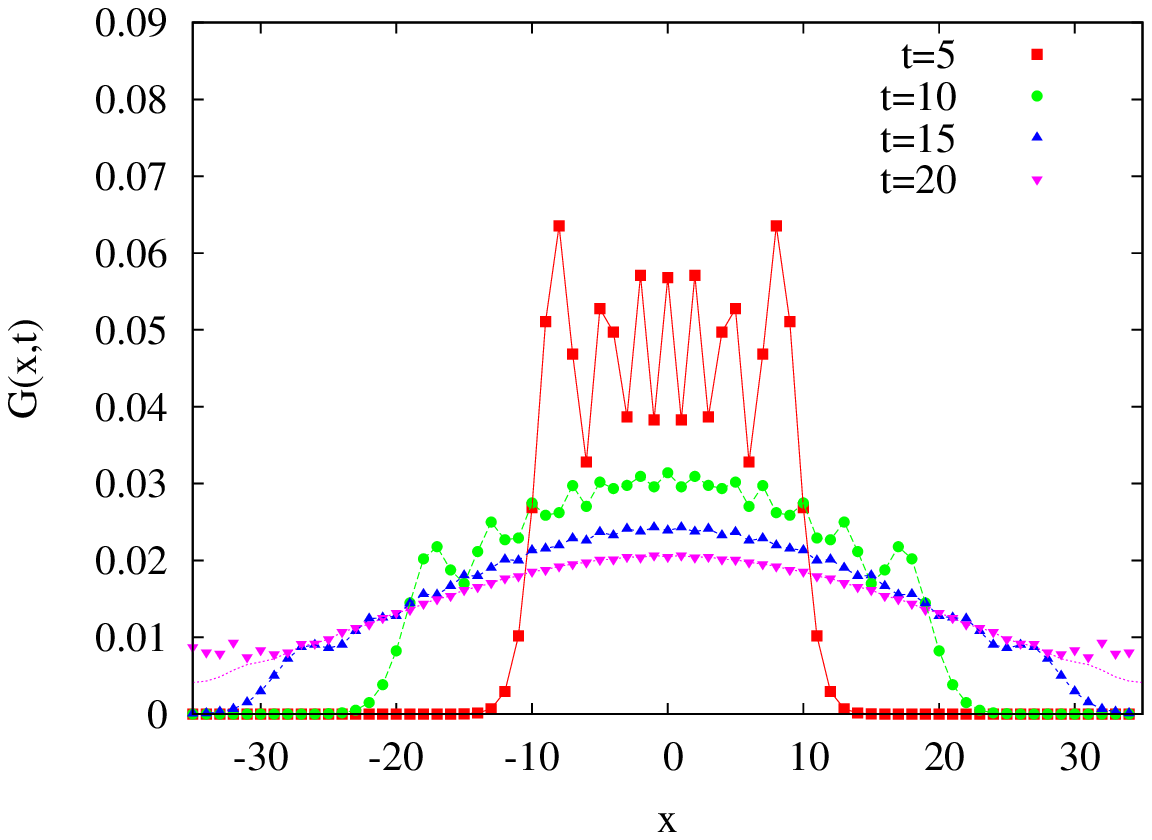}
\includegraphics[scale=0.6]{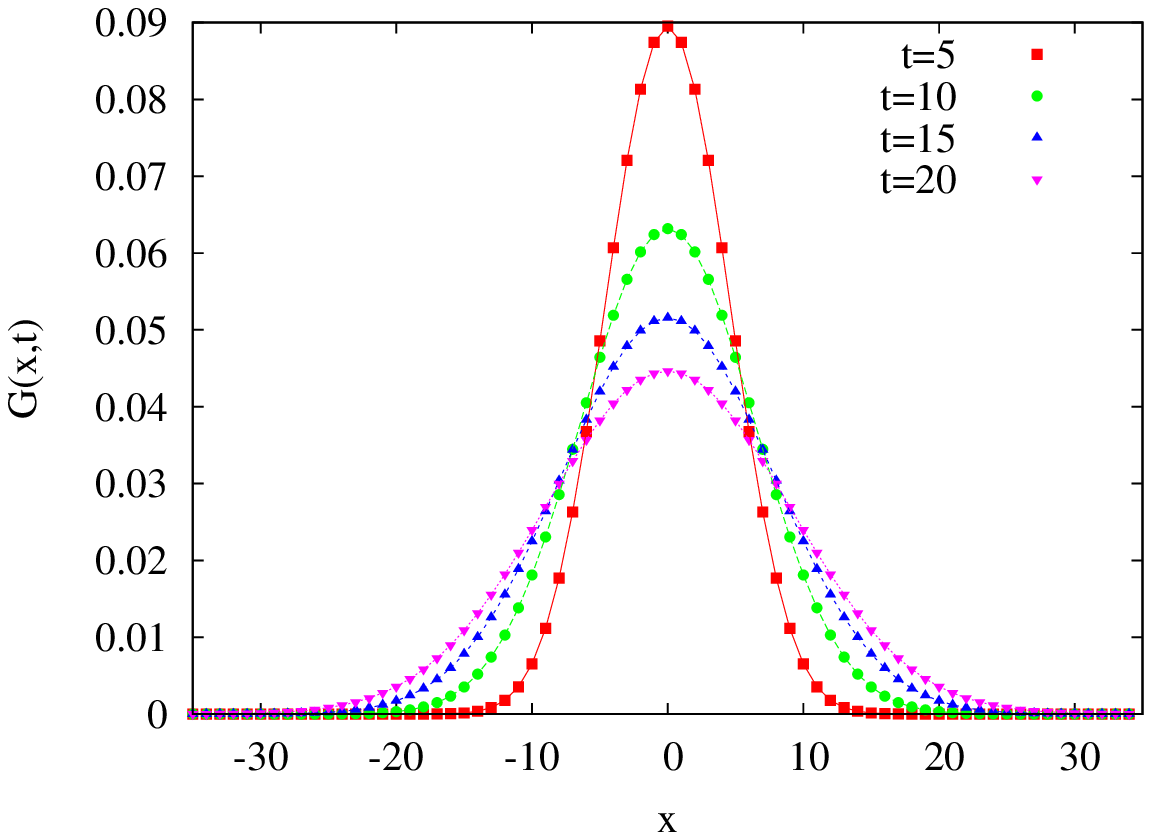}
\caption{Green's function of the density for different times and parameter values $\gamma=0.1$ (left)
and $\gamma = 1$ (right). The tunneling rate is set to $\lambda = 1$. Points represent the exact solution
for finite $N=70$ by evaluating the sums in (\ref{eq:gfsum}).}
\label{fig:gfx}
\end{figure}

Finally, we compare the exact numerical results of the density evolution obtained for a chain of finite 
length $N=70$ with the numerically evaluated integrals for $\mathcal{G}(x,t)$. As shown in Fig. \ref{fig:gfx}, the
data show excellent agreement and verifies the analytical result for the Green's function. On the left hand
side a relatively small stochastic hopping rate $\gamma =0.1$ allows the ballistic features to survive for short times,
but for larger times the crossover to a Gaussian profile can be observed. The discrepancy of the finite size results
for $t=20$ is a simple consequence of the ring geometry, since the two ends of the expanding wavefront collide.
The right hand side shows results for $\gamma = 1$, where already the smallest time shown is in the diffusive regime.
The difference between the speeds of spreading is clearly visible.

%

\section{Discussion\label{sec:disc}}

We have presented an exact analytical solution for the Green's function of the density profile in the symmetric
quantum exclusion process. The derivation is based on a more general framework for the treatment of the
quantum master equation for stochastic processes with quadratic and hermitian Lindblad generators.
In our specific model the Liouvillian becomes analogous to an operator which describes a coupled system of
two species of fermions with nearest-neighbour interactions. This form allowed an exact treatment of the
one- and two-point correlation functions.

For the complete solution of the problem one should also look at the higher order correlations.
The form of the Liouvillian (\ref{eq:liouqep}) suggests that the solution should be available by Bethe ansatz.
This would generate the $n$-point functions in a form similar to (\ref{eq:bag}) and (\ref{eq:baf}) but with
additional terms where all the different permutations of the phase factors are present. The corresponding
scattering phases should then factorize into pair-terms, giving a factor $S_{ji}$ when the permutation exchanges
a particle at position $i$ with a hole at position $j$ and a factor a $\tilde S_{ji}$ when two particle-operators are exchanged.
Although this scheme looks feasible, a rigorous proof such as the one recently given for the simple exclusion process
\cite{TW08} would be desirable.

It is important to stress that the short and long time behaviour of the density results from
the asymptotic behaviour of the Green's function for large and small wavelengths, respectively.
Therefore, it can be interpreted as a competition between ballistic and diffusive channels for the spectral weight 
in the propagator. This is exactly the same mechanism which was outlined in \cite{Sirker11} by the discussion
of the thermal spin-spin correlation function for the XXZ chain. Note, that in our case
the large wavenumbers are always exponentially suppressed for long times and therefore the diffusive
transport channel dominates.

With the solution for the Green's function in hand, one could further look at evolutions from some simple initial
conditions, such as a step function in the density profile. This situation was investigated recently for the symmetric simple
exclusion process \cite{Derrida09} as well as in the purely coherent case \cite{ARRS99,Antal08}. Apart from the crossover in the
evolution of densities, one could look at the fluctuations of the particle number where an interesting crossover
from $\sqrt{t}$ \cite{Derrida09} to $\log t$ scaling \cite{Antal08} should emerge.

Although the problem was formulated in the language of fermions, it is straightforward to generalize it to
certain spin models, such as the XY chain, which can be transformed into quadratic fermionic Hamiltonians.
However, one also has to fulfill the requirement that the stochastic terms translate into quadratic and hermitian
Lindblad operators. The simplest example is the so-called dephasing noise, which is described by a $\sigma_z$
term and is therefore included in this class.

Another extension of the present work could be considering different boundary conditions.
It would be interesting to check whether the problem remains solvable if one has, instead of a ring,
a linear chain with boundary injection and ejection at the ends. Since an exact solution for the
steady state has been recently presented in case of the XX model with dephasing noise 
\cite{Znidaric10,Znidaric11},
one could speculate whether our framework could generalize results considering the entire dynamics.

The treatment of the two-point correlation functions in the present many-body problem greatly parallels
to the single-particle quantum diffusion problem of Esposito and Gaspard \cite{EG05}.
The only difference is in the actual form of the function $\beta$ which, however, does not change
the qualitative picture and, in the thermodynamic limit, leads to diffusion on large timescales.
We expect that this behaviour would not change by considering a stochastic process with longer range
hopping, although the effective diffusion constant might increase. Therefore it remains a puzzling question,
whether one could construct some more complicated stochastic processes which would eventually protect
the ballistic features and lead to the onset of diffusion at a \emph{finite} value of the corresponding rate.

Finally, one should point out that the hermiticity of the Lindblad operators was found to be a sufficient
condition for the existence of the correlation function hierarchy. It would be interesting to address the
question whether it is also a necessary one. Furthermore, one could investigate whether the inclusion of
some interaction term in the Hamiltonian would still leave the hierarchy unchanged.

\ack The author would like to thank Ingo Peschel and Michael Wolf for discussions.
He acknowledges financial support by the Danish Research Council, QUANTOP
and the EU projects COQUIT, QUEVADIS and QUERG.

\appendix

\section{Time evolution of the $n$-point correlations}

In this appendix we derive the formulas that are used to obtain the time evolution
of the $n$-point functions $\Gamma_{\underline{\nu}}$ introduced in Eq. (\ref{eq:npf}).
Using the commutation relation $\{ c_k , c_l \} =2 \delta_{k,l}$ it is easy to show that
\eq{
\left[c_k c_l,\Gamma_{\underline{\nu}} \right] = 
2\left( \hat{c}^\dag_k\hat{c}_l - \hat{c}^\dag_l\hat{c}_k \right)
\Gamma_{\underline{\nu}}
\label{eq:commut}}
where the superoperators $\hat{c}^\dag_k$ and $\hat{c}_k$ that insert or remove
a Majorana operator from the string $\Gamma_{\underline{\nu}}$ were defined in (\ref{eq:so}).
Together with the definition of the Hamiltonian in Eq. (\ref{eq:hlmatrix})
and $H_{kl}=-H_{lk}$ this yields
\eq{
\mathcal{L}_{\mathrm{coh}}^\dag(\Gamma_{\underline{\nu}}) =
i\left[H, \Gamma_{\underline{\nu}} \right] = 
-\sum_{k,l=1}^{2N} H_{kl} \hat{c}^\dag_k\hat{c}_l \Gamma_{\underline{\nu}}
\label{eq:cohev}}
and one obtains the coherent part of the evolution. Setting
$\mathcal{L} = \mathcal{L}_{\mathrm{coh}} + \sum_\alpha \mathcal{L_\alpha}$,
one has to calculate the contribution
\eq{
\mathcal{L_\alpha^\dag}(\Gamma_{\underline{\nu}}) = \frac{1}{32}
\sum_{ijkl}L_{\alpha,ij}L_{\alpha,kl}
\left(\left\{c_i c_j c_k c_l , \Gamma_{\underline{\nu}} \right\} - 2c_i c_j \Gamma_{\underline{\nu}} c_k c_l \right)
}
where we used the hermiticity $L_\alpha^\dag=L_\alpha$ of the Lindblad operators.
The expression in the parenthesis can be rewritten as
$c_i c_j \left[c_k c_l,\Gamma_{\underline{\nu}}\right] - \left[c_i c_j,\Gamma_{\underline{\nu}}\right]c_k c_l$
and using the symmetry under exchanging the pair of indices
$(i,j) \leftrightarrow (k,l)$ one finds
\eq{
\mathcal{L_\alpha^\dag}(\Gamma_{\underline{\nu}}) =
\frac{1}{32}\sum_{ijkl}L_{\alpha,ij}L_{\alpha,kl} 
\left[c_i c_j ,\left[c_k c_l,\Gamma_{\underline{\nu}}\right]\right] \, .
}
Finally, applying (\ref{eq:commut}) twice and using $L_{\alpha,kl}=-L_{\alpha,lk}$ one arrives at
\eq{
\mathcal{L_\alpha^\dag}(\Gamma_{\underline{\nu}}) = 
\frac{1}{2}\sum_{ijkl}L_{\alpha,ij}L_{\alpha,kl} 
\hat{c}^\dag_i\hat{c}_j \hat{c}^\dag_k\hat{c}_l \Gamma_{\underline{\nu}} \, .
}
The full time evolution operator (\ref{eq:liou}) is then obtained by normal ordering, summing over $\alpha$
and adding the coherent contribution in Eq. (\ref{eq:cohev}).

\section{Solution of the Bethe equation}

The quantization of the wavenumber $\theta$ is determined by Eqs. (\ref{eq:gl}) with indices
$l=0$ and $l=N-1$. The system of these two equations for the amplitudes $A$ and $B$ in 
(\ref{eq:scp}) has a nontrivial solution only if the determinant of the coefficients vanishes.
This condition leads to the following Bethe equation
\eq{
i\beta \sin \theta\left[\cos \theta N - R(q)\right]=\sin \theta N
\label{eq:beq}}
where the parameters are defined as
\eq{
\beta = 2 \frac{\lambda}{\gamma}\frac{\sin q/2}{\cos q}, \qquad
R(q) = \frac{1}{2}\left( i^N e^{-iN\frac{q}{2}} + i^{-N} e^{iN\frac{q}{2}}\right).
\label{eq:beta}}
The second equation then fixes the ratio of the amplitudes $B/A=S_{21}$.
Note, that the form of $R(q)$ is set solely by the boundary condition. 
\par
The solution of the Bethe equation requires some attention since it depends on the parity of $N$
as well as $q_j$. The problem is essentially the same as the one treated in \cite{EG05} only the
exact form of the parameter $\beta$ differs. This, however, does not change the qualitative
form of the spectrum and leads to the same families of eigenvalues.
In the following we give a short summary of the results on the spectrum and refer to \cite{EG05}
for a detailed analysis. For the sake of concreteness we choose $N=4k+2$ with an arbitrary integer $k$.
\subsection*{The complex eigenvalues $\theta_I$}
The first family is obtained by solving
\eq{
i\beta \sin \theta = \left\{
\begin{array}{rr}
-\cot \frac{N \theta}{2} & q_j \mbox{ odd} \\
\tan\frac{N \theta}{2} & q_j \mbox{ even}
\end{array}
\right.
\label{eq:beq1}}
and the scattering phase is given by
\eq{
S_{21} = -\frac{1-\xi}{1+\xi}, \qquad \xi = \beta\sin{\theta}.
\label{eq:scp1}}
In general, Eq. (\ref{eq:beq1}) yields complex solutions for $\theta$ which
have different asymptotic expressions depending on the value of $\xi$.
For $|\xi| < 1$ one has $\theta_< = \theta_{0<}+i\delta_{<}$ where the real
and imaginary parts read
\eq{
\theta_{0<} = \left\{
\begin{array}{ll}
\frac{\pi}{N}(2n+1), & n=0,1,\dots,\frac{N}{2}-1 \\
\frac{\pi}{N}2n, & n=1,\dots,\frac{N}{2}-1
\end{array}
\right. , \quad
\delta_< = \frac{1}{N}\log\frac{1+\xi_<}{1-\xi_<}
\label{eq:th1<}}
with $\xi_< = \beta \sin \theta_{0<}$. The upper and the lower solutions refer again to $q_j$ being odd
and even, respectively. For $|\xi|>1$ the solutions $\theta_> = \theta_{0>}+i\delta_{>}$ are given by
\eq{
\theta_{0>} = \left\{
\begin{array}{ll}
\frac{\pi}{N}2n, & n=1,\dots,\frac{N}{2}-1 \\
\frac{\pi}{N}(2n+1), & n=0,1,\dots,\frac{N}{2}-1
\end{array}
\right. , \quad
\delta_> = \frac{1}{N}\log\frac{\xi_>+1}{\xi_>-1}
\label{eq:th1>}}
with $\xi_> = \beta \sin \theta_{0>}$. Because of the condition $|\xi| > 1$, the solutions $\theta_>$
only exist in the interval $\theta_c < \theta_0 < \pi - \theta_c$ with $\theta_c = \arcsin |\beta|^{-1}$.
In case $\theta_c < \pi/N$, two special solutions for $q_j$ odd appear at $\theta = \delta$
and $\theta = \pi - \delta$ with $\delta^2=\frac{2i}{\beta N}$.
Note, that the corrections to these asymptotic forms are $\mathcal{O}(1/N^2)$ except
from the vicinity of the branch cut $|\xi | \approx 1$ where the approximation is not valid.
\subsection*{The real eigenvalues $\theta_{II}$}

Equation (\ref{eq:beq}) can also be satisfied by setting $\cos N\theta = R(q)=\pm 1$
where the $\pm$ sign refers to $q_j$ being odd and even, respectively. The solutions are
\eq{
\theta_{II} = \left\{
\begin{array}{ll}
\frac{\pi}{N}2n, & n=1,\dots,\frac{N}{2}-1 \\
\frac{\pi}{N}(2n+1), & n=0,1,\dots,\frac{N}{2}-1
\end{array}
\right. .
\label{eq:th2}}
The scattering phase is exactly $S_{21}=-1$ and therefore this family constitutes the antisymmetric
solutions of the potential scattering problem. The latter property is also evident from the fact that
the solutions are independent of $\beta$ and thus of the exact form of the potential.

\subsection*{The diffusive eigenvalues $\theta_{d}$}

There exists a special family of eigenvalues which in turn correspond to
the bound state solutions of the Bethe equation. Setting $\theta_d = \pm \frac{\pi}{2} + i\eta$
one has
\eq{
|\beta| \cosh \eta = \left\{
\begin{array}{ll}
\tanh \frac{N \eta}{2} & q_j \mbox{ odd}\\
\coth \frac{N \eta}{2} & q_j \mbox{ even}
\end{array}
\right.
\label{eq:beq3}}
where the sign $\pm$ refers to the cases $\beta>0$ and $\beta<0$, respectively.
In the $N \to \infty$ limit the solutions are given by
\eq{
\eta = \acosh |\beta|^{-1}
\label{eq:scp3}}
and, except from the vicinity $|\beta| \approx 1$, finite size corrections $\delta \eta$ are 
exponentially small in $N$
\eq{
\delta \eta \approx 2 \frac{e^{-N\acosh |\beta|^{-1}}}{\sqrt{1-\beta^2}}.
\label{eq:deta}}
Therefore $\xi \to 1$ and the scattering phase vanishes, corresponding to exponentially decaying, localized eigenmodes.

\begin{figure}[htb]
\center
\includegraphics[scale=0.65]{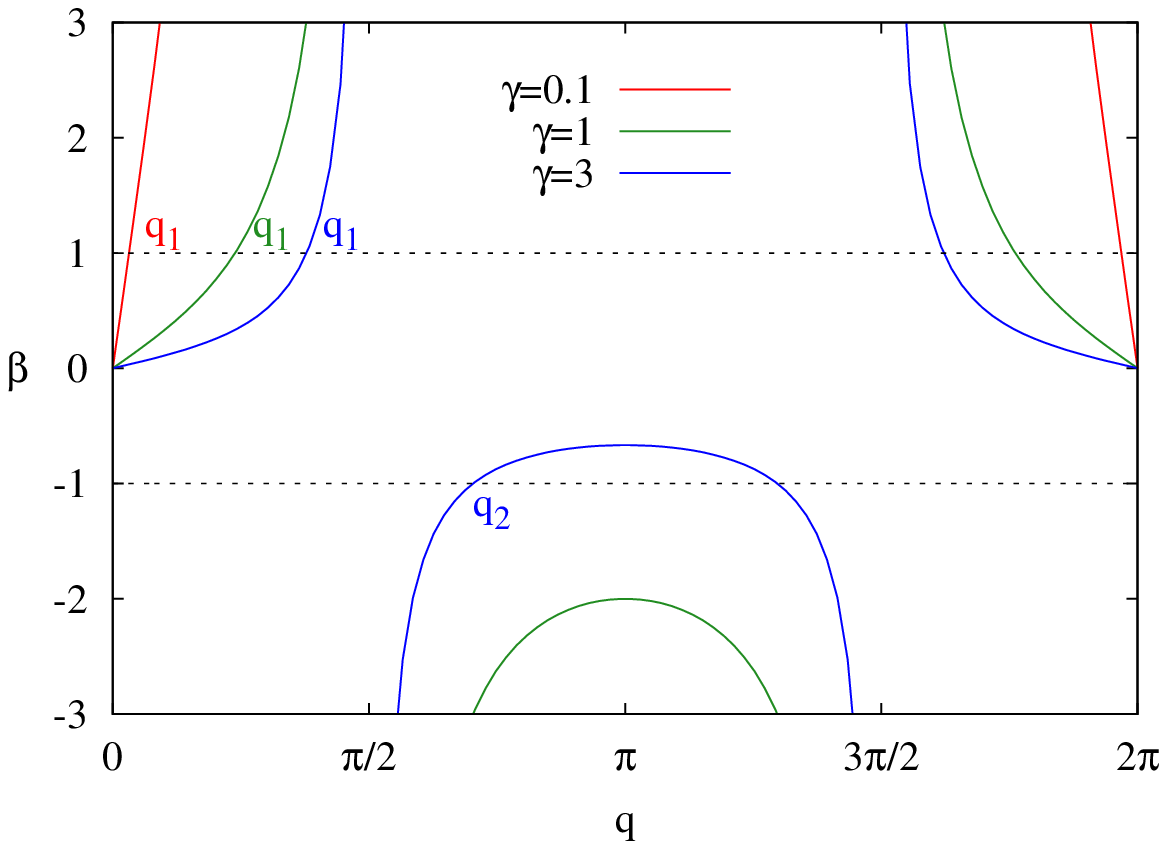}
\caption{Plots of the function $\beta$ in Eq. (\ref{eq:beta}) for different values of $\gamma$
and $\lambda=1$. The solutions $q_1$ of $\beta=1$ are indicated in corresponding color.
For $\gamma=3$ the wavenumber $q_2$ indicates the solution $\beta=-1$.}
\label{fig:betaq}
\end{figure}

The diffusive eigenvalues have to satisfy $|\beta| < 1$. Thus, for $\gamma<2\lambda$, they only exist in the
intervals $0 \le q < q_1$ and $2\pi-q_1 < q < 2\pi$ where $q_1$ is determined by the condition $\beta(q_1)=1$.
For larger values of $\gamma$ one has an additional interval $q_2< q < 2\pi-q_2$ where $\beta(q_2)=-1$.
The above conditions are sketched in Fig. \ref{fig:betaq} for the same values of $\gamma$ as used in
Fig. \ref{fig:z1z2}, see text. For $\gamma \gg \lambda$ the wavenumber $q_1 \to \pi/2$ from below as well as
$q_2 \to \pi/2$ from above and one has $\theta_d$ eigenvalues almost everywhere in the spectrum, recovering
purely diffusive behaviour.

%

\section*{References}

\bibliographystyle{iopart-num}

\bibliography{sqep_refs}

\providecommand{\newblock}{}
\begin{thebibliography}{10}
\expandafter\ifx\csname url\endcsname\relax
  \def\url#1{{\tt #1}}\fi
\expandafter\ifx\csname urlprefix\endcsname\relax\def\urlprefix{URL }\fi
\providecommand{\eprint}[2][]{\url{#2}}

\bibitem{Giamarchibook}
Giamarchi T 2004 {\em Quantum Physics in One Dimension\/} (Clarendon Press,
  Oxford)

\bibitem{Castella95}
Castella H, Zotos X and Prelovsek P 1995 Integrability and ideal conductance at
  finite temperatures {\em Phys. Rev. Lett.\/} {\bf 74} 972

\bibitem{Sirker09}
Sirker J, Pereira R~G and Affleck I 2009 Diffusion and ballistic transport in
  one-dimensional quantum systems {\em Phys. Rev. Lett.\/} {\bf 103} 216602

\bibitem{Sirker11}
Sirker J, Pereira R~G and Affleck I 2011 Conservation laws, integrability and
  transport in one-dimensional quantum systems {\em Phys. Rev. B\/} {\bf 83}
  035115

\bibitem{HHB07}
For a review see: Heidrich-Meisner F, Honecker A and Brenig W 2007 Transport in quasi
  one-dimensional spin-1/2 systems {\em Eur. Phys. J. Special Topics\/} {\bf
  151} 135

\bibitem{BPbook}
Breuer H~P and Petruccione F 2002 {\em The Theory of Open Quantum Systems\/}
  (Oxford Univ. Press, Oxford)

\bibitem{Wichterich07}
Wichterich H, Henrich M~J, Breuer H~P, Gemmer J and Michel M 2007 Modeling heat
  transport through completely positive maps {\em Phys. Rev. E\/} {\bf 76}
  031115

\bibitem{Prosen08}
Prosen T 2008 Third quantization: a general method to solve master equations
  for quadratic open fermi systems {\em New J. Phys.\/} {\bf 10} 043026

\bibitem{KP09}
Karevski D and Platini T 2009 Quantum non-equilibrium steady states induced by
  repeated interactions {\em Phys. Rev. Lett.\/} {\bf 102} 207207

\bibitem{Prosen11}
Prosen T 2011 Open xxz spin chain: Nonequilibrium steady state and strict bound
  on ballistic transport (\textit{Preprint} \eprint{1103.1350})

\bibitem{Znidaric10}
\v{Z}nidari\v{c} M 2010 Exact solution for a diffusive nonequilibrium steady state of
  an open quantum chain {\em J. Stat. Mech.\/} p L05002

\bibitem{Znidaric09}
\v{Z}nidari\v{c} M 2010 Dephasing-induced diffusive transport in the anisotropic
  heisenberg model {\em New J. Phys.\/} {\bf 12} 043001

\bibitem{TWV09}
Temme K, Wolf M~M and Verstraete F 2009 Stochastic exclusion processes versus
  coherent transport (\textit{Preprint} \eprint{0912.0858})

\bibitem{Clark09}
Clark S~R, Prior J, Hartmann M~J, Jaksch D and Plenio M~B 2009 Exact matrix
  product solutions in the heisenberg picture of an open quantum spin chain
  {\em New J. Phys.\/} {\bf 12} 025005

\bibitem{PZ09}
Prosen T and \v{Z}nidari\v{c} M 2009 Matrix product simulations of non-equilibrium
  steady states of quantum spin chains {\em J. Stat. Mech.\/} p P02035

\bibitem{EG05}
Esposito M and Gaspard P 2005 Exactly solvable model of quantum diffusion {\em
  J. Stat. Phys.\/} {\bf 121} 463

\bibitem{EG05prb}
Esposito M and Gaspard P 2005 Emergence of diffusion in finite quantum systems
  {\em Phys. Rev. B\/} {\bf 71} 214302

\bibitem{Mohseni08}
Mohseni M, Rebentrost P, Lloyd S and Aspuru-Guzik A 2008 Environment-assisted
  quantum walks in photosynthetic energy transfer {\em Journal of Chemical
  Physics\/} {\bf 129} 174106

\bibitem{Plenio08}
Plenio M and Huelga S 2008 Dephasing assisted transport: Quantum networks and
  biomolecules {\em New J. Phys.\/} {\bf 10} 113019

\bibitem{Liggettbook}
Liggett T 1999 {\em Stochastic Interacting Systems: Contact, Voter and
  Exclusion Processes\/} (Springer Berlin)

\bibitem{Lindblad76}
Lindblad G 1976 On the generators of quantum dynamical semigroups {\em Comm.
  Math. Phys.\/} {\bf 48} 119

\bibitem{Znidaric11}
\v{Z}nidari\v{c} M 2011 Solvable quantum nonequilibrium model exhibiting a phase
  transition and a matrix product representation {\em Phys. Rev. E\/} {\bf 83}
  011108

\bibitem{Schutz97}
Sch{\"u}tz G~M 1997 Exact solution of the master equation for the asymmetric
  exclusion process {\em J. Stat. Phys.\/} {\bf 88} 427

\bibitem{GRbook}
Gradshteyn I~S and Ryzhik I~M 1980 {\em Table of Integrals, Series, and
  Products\/} (Academic Press, New York)

\bibitem{TW08}
Tracy C and Widom H 2008 Integral formulas for the asymmetric simple exclusion
  process {\em Communications in Mathematical Physics\/} {\bf 279} 815

\bibitem{Derrida09}
Derrida B and Gerschenfeld A 2009 Current fluctuations of the one dimensional
  symmetric simple exclusion process with step initial condition {\em J. Stat.
  Phys.\/} {\bf 136} 1

\bibitem{ARRS99}
Antal T, R{\'a}cz Z, R{\'a}kos A and Sch{\"u}tz G~M 1999 Transport in the xx
  chain at zero temperature: Emergence of flat magnetization profiles {\em
  Phys. Rev. E\/} {\bf 59} 4912

\bibitem{Antal08}
Antal T, Krapivsky P~L and R{\'a}kos A 2008 Logarithmic current fluctuations in
  non-equilibrium quantum spin chains {\em Phys. Rev. E\/} {\bf 78} 061115

\end{thebibliography}

\end{document}